\documentclass[aps,pra,superscriptaddress,twocolumn,10pt]{revtex4-2}

\usepackage{amsmath}
\usepackage{amssymb}
\usepackage{amsthm}
\usepackage{bbm}
\usepackage{color}
\usepackage{graphicx}[floatfix]
\usepackage{hyperref}
\usepackage[utf8]{inputenc}
\usepackage[T1]{fontenc}
\usepackage{physics}
\usepackage{times}
\usepackage{ulem}
\usepackage{xcolor}

\hypersetup{
    colorlinks=true,linkcolor=blue,citecolor=blue,
    filecolor=blue,urlcolor=blue,breaklinks=true
}

\newcommand{\orcid}[1]{\href{https://orcid.org/#1}
{\includegraphics[width=7pt]{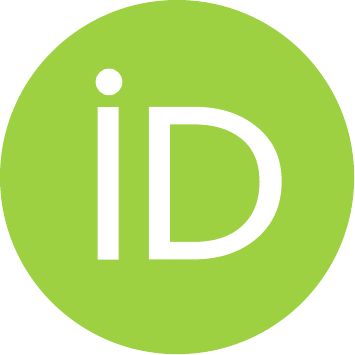}}}

\theoremstyle{definition}

\def\be{\begin{equation}}
\def\ee{\end{equation}}

\def\bc{\begin{center}}
\def\ec{\end{center}}
\def\bal{\begin{align}}
\def\eal{\end{align}}

\begin{document}

\title{Entanglement between uncoupled modes with time-dependent complex frequencies}

    \author{D. Cius\orcid{0000-0002-4177-1237}}
    \email{danilocius@gmail.com}
    \affiliation{
        Departamento de F\'{i}sica Matem\'{a}tica, Instituto de F\'{i}sica da Universidade de S\~{a}o Paulo, 05508-090 S\~{a}o Paulo, Brazil
}

        \author{G. M. Uhdre\orcid{0000-0003-0344-8532}}
        \affiliation{
                Programa de P\'os-Gradua\c{c}\~{a}o em Ci\^{e}ncias/F\'{i}sica,
                Universidade Estadual de Ponta Grossa,
                84030-900 Ponta Grossa, Paran\'a, Brazil
        }

        \author{A. S. M. de Castro\orcid{0000-0002-1521-9342}}
        \affiliation{
            Programa de P\'os-Gradua\c{c}\~{a}o em Ci\^{e}ncias/F\'{i}sica,
                Universidade Estadual de Ponta Grossa,
                84030-900 Ponta Grossa, Paran\'a, Brazil
        }
        \affiliation{
                Departamento de F\'{i}sica,
                Universidade Estadual de Ponta Grossa,
                84030-900 Ponta Grossa, Paran\'{a}, Brazil
        }

        \author{Fabiano M. Andrade\orcid{0000-0001-5383-6168}}
        \affiliation{
                Programa de P\'os-Gradua\c{c}\~{a}o em Ci\^{e}ncias/F\'{i}sica,
                Universidade Estadual de Ponta Grossa,
                84030-900 Ponta Grossa, Paran\'a, Brazil
        }
        \affiliation{
          Departamento de Matem\'{a}tica e Estat\'{i}stica,
          Universidade Estadual de Ponta Grossa,
          84030-900 Ponta Grossa, Paran\'{a}, Brazil
        }

\date{\today}

\begin{abstract}
In this work we present the general unified description for the unitary time-evolution generated by time-dependent non-Hermitian Hamiltonians embedding the bosonic representations of $\mathfrak{su}(1,1)$ and $\mathfrak{su}(2)$ Lie algebras. We take into account a time-dependent Hermitian Dyson maps written in terms of the elements of those algebras with the relation between non-Hermitian and its Hermitian counterpart being independent of the algebra realization. 
As a direct consequence, we verify that a time-evolved state of uncoupled modes modulated by a time-dependent complex frequency may exhibits a non-zero entanglement even when the cross-operators, typical of the interaction between modes, are absent. This is due the non-local nature of the non-trivial dynamical Hilbert space metric encoded in the time-dependent parameters of the general Hermitian Dyson map, which depend on the imaginary part of the complex frequency. We illustrate our approach by setting the $\mathcal{PT}$-symmetric case where the imaginary part of frequency is linear on time for the two-mode bosonic realization of Lie algebras.

\end{abstract}


\maketitle

\section{Introduction}
\label{sec:introduction}

Entanglement as a quantum resource can naturally arise when more than one quantum system is considered, and therefore has become an indispensable tool for characterizing  quantum many-body systems \cite{eisert2010}. It is essential for many applications in the field of quantum information, such as quantum key distribution \cite{ekert1992}, quantum teleportation \cite{bennett1993} and quantum computation \cite{raussendorf2001}. The generation of entanglement in optical setups may be achieved, for example, by means of parametric amplifiers considered for cavities modes whose coupling dynamics is modeled by a bilinear Hamiltonian $\hat{H}\propto \hat{a}_{1}\hat{a}_{2} + \text{H.c}$ \cite{heidmann1987,zhang2017}. Also, beam splitter acts as an entangler, and is described by the Hamiltonian $\hat{H}\propto \hat{a}_{1}\hat{a}_{2}^\dagger + \text{H.c}$ \cite{paris1999,kim2002}. Nevertheless, entanglement may also be produced and manipulated in different physical scenarios as NMR \cite{singh2018}, superconducting qubits \cite{egger2019}, optomechanical systems \cite{dixon2020} and others \cite{rauschenbeutel2001,yang2004}. The entanglement characterization is based on well-established criteria which provide various conditions and require different experimental techniques to be evaluated \cite{horodecki2009,guhne2009}. As an example, in continuum-variables systems, the Hillery and Zubairy criteria \cite{hillery2006PRL,hillery2006PRA} and the Nha and Zubairy criterium \cite{nha2008,nha2007} provide a class of inequalities involving the two-mode bosonic realizations of $\mathfrak{su}(1,1)$ and $\mathfrak{su}(2)$ Lie algebras whose violation indicates the presence of entanglement. Some conditions for entanglement in multipartite systems are derived in Ref. \cite{hillery2010}. Genuine multipartite entanglement and EPR steering for continuum-variables are discussed by Teh and Reid \cite{teh2014}, which may be applied on three-mode spontaneous parametric down-conversion to detect tripartite genuine non-gaussian entanglement \cite{agusti2020}. In this sense, we extend the study of entanglement quantum systems considered in the scope of non-hermiticity.

Non-Hermitian systems can be used to model open and dissipative dynamics assuming the dynamics described by means of non-Hermitian Hamiltonian operators as found in Refs. \cite{gisin1981,sergi2013,michishita2020,cius2022PRA}. However, new perspectives have been addressed to the foundations of quantum theory since the publication of the seminal work of Bender and Boettcher \cite{bender1998}. They conjectured that a non-Hermitian Hamiltonian have real eigenvalues if it exhibits unbroken parity-time ($\mathcal{PT}$) symmetry \cite{bender2005}. Shortly after, a formalism embracing the special class of non-Hermitian Hamiltonians, called pseudo-Hermitian, was presented by Mostafazadeh \cite{mostafazadeh2002a,mostafazadeh2002b,mostafazadeh2002c}. He showed that pseudo-hermiticity and the existence of a positive-defined metric of Hilbert space are the necessary conditions for an operator to exhibit a real spectrum, and generate a unitary time-evolution within a consistent quantum mechanical 
framework. 
Simulations of quantum $\mathcal{PT}$-symmetric systems are usually done in optical experimental setups, in the sense that the mathematical equivalence between the quantum mechanical Schrödinger equation and the optical wave equation allows the realization of complex potentials within the framework of optics as suggested in Refs. \cite{guo2009,makris2008,makris2010,chen2017}. Similar simulations of non-Hermitian quantum systems are found in the context of silicon photonics \cite{wang2021}. In addition, it is noteworthy the relevance of these mathematical developments in many topics of physics as complex scattering potentials \cite{ahmed2001,mostafazadeh2009PRL,mostafazadeh2013}, tight-binding chain \cite{jin2009}, anisotropic $XY$ model \cite{zhang2013}, quantum brachistochrone problem in both theoretical \cite{bender2007,gunter2008} and experimental \cite{zheng2013} scenarios, coupled optomechanical systems \cite{xu2015}, geometric phase \cite{maamache2015}, pseudochirality \cite{rivero2020}, and non-Hermitian version of Jaynes-Cummings optical model obtained from $\kappa$-deformed Dirac oscillator \cite{uhdre2022}. Furthermore, in the field of quantum information, there are many interesting investigations in the context of pseudo-hermiticity considered for optimal-speed evolution generation \cite{mostafazadeh2009PRA},  pseudo-Hermitian networks \cite{jin2011}, perfect state transfer in non-Hermitian networks \cite{zhang2012},  information retrieval \cite{kawabata2017},  holonomic gates \cite{pinske2019}, and an experimental quantum cloning protocol was presented in Ref. \cite{zhan2020}. Recently, an efficient simulation scheme of a finite $\mathcal{PT}$-symmetric system with LOCC was proposed in Ref. \cite{li2022}.

In addition, when a time-dependent non-Hermitian Hamiltonian is considered, Mostafazadeh demonstrated that by applying a time-dependent metric operator we cannot ensure the unitarity of the dynamics simultaneously with the observability of the Hamiltonian as happens in the time-independent scenario \cite{mostafazadeh2007,mostafazadeh2018}. To settle this issue, Fring and Moussa \cite{fring2016a} have proposed to treat the Hamiltonian as a mere generator of time-evolution preserving the unitary dynamics and assuming a time-dependent metric which leads to a distinction between the observable energy operator and the Hamiltonian \cite{fring2017}. Recently, it has been shown that these time-dependent non-Hermitian Hamiltonians can generate squeezed states of the radiation field with an infinite degree of squeezing in a finite time interval \cite{deponte2019,deponte2021E}, what is also can be achieved by considering a single-mode oscillator modulated by a time-dependent pure imaginary frequency in applying a quadratic non-Hermitian Dyson map \cite{dourado2021}. These previous results also lead to an enhancement in the photon production by the Dynamical Casimir Effect (DCE) described by a DCE-like non-Hermitian Hamiltonian \cite{cius2022PA}.

In time-dependent pseudo-Hermitian context, the no-go theorems are discussed in Ref. \cite{ju2019} in a modified formulation of pseudo-Hermitian quantum mechanics based on the geometry of Hilbert spaces \cite{ju2022}.  Moreover, Fring and Frith \cite{fring2019} investigated the von Neumann entropy behavior of an interacting 
$\mathcal{PT}$-symmetric bosonic system, and they verify that the entropy decays asymptotically to a finite constant value when the symmetry is spontaneously broken. Frith \cite{frith2020} also considers a non-Hermitian version of the Jaynes-Cummings model in which an exotic behavior of the entanglement appears in the  $\mathcal{PT}$-symmetric broken regime.

In this work, we consider a more general Hermitian counterpart embedding both $\mathfrak{su}(1,1)$ and $\mathfrak{su}(2)$ Lie algebras as done in \cite{maamache2017,koussa2018}.
We observe that a pseudo-Hermitian Hamiltonian describing uncoupled modes, modulated by a time-dependent complex function, can exhibit entanglement even when the cross-operators, typical of the interaction between modes, are absent. We evaluate the entanglement between two modes by means of the linear entropy \cite{horodecki2009}. 
The non-trivial entanglement between uncoupled modes comes from the general time-dependent Hermitian Dyson map, which generates a dynamical inseparable Hilbert space through which one of the quantum correlations are encoded.
In this context, it is significant to understand the relevance of the physics following from pseudo-Hermitian Hamiltonians and its possible applications to quantum information.

This manuscript is organized as follows:
Section \ref{sec:TDNHS} brings a brief review of the formalism of time-dependent pseudo-Hermitian Hamiltonian with time-dependent metric, showing how to build the time-dependent Dyson map to obtain the general Hermitian counterpart Hamiltonian carrying in a unified form both $\mathfrak{su}(1,1)$ and $\mathfrak{su}(2)$ Lie algebras. Then, we exactly solve the time-dependent Schrödinger equation to have a unitary time-evolution operator. In Section \ref{sec:K0}, we consider the simplest pure algebraic case, which can be analyzed exactly. We present the usual two-mode bosonic realizations of $\mathfrak{su}(1,1)$ and $\mathfrak{su}(2)$ in Section \ref{sec:entanglement}, we discuss the cases where we achieve the maximum entanglement between the modes by means of the linear entropy. We also present a qualitative discussion toward a multimode bosonic realization for our approach. Finally, in Section \ref{sec:conclusion}, we point out our main conclusions.

\section{Time-Dependent Non-Hermitian System}
\label{sec:TDNHS}

\subsection{Non-Hermitian approach}


It has been shown that a time-dependent non-Hermitian Hamiltonian operator can  generate a unitary time-evolution provided that there are the following two Schrödinger equations ($\hbar=1$) \cite{fring2016a}
\begin{subequations}
        \label{TDSE}
        \begin{align}
        \label{TDSE-H}
        i\partial_{t}\vert \Psi (t) \rangle &= \hat{H}(t)\vert \Psi (t) \rangle, \qquad \hat{H}(t)\neq \hat{H}^{\dagger}(t),
        \\
        \label{TDSE-h}
        i\partial_{t}\vert \psi (t) \rangle &= \,\hat{h}(t)\vert \psi (t) \rangle, \qquad ~\,\, \hat{h}(t) = \hat{h}^{\dagger}(t),
        \end{align}
\end{subequations}
with both states related by means of the time-dependent Dyson map $\hat{\eta}(t)$ defined as
\begin{equation}
\label{psiPsi}
\vert \psi (t) \rangle = \hat{\eta}(t) \vert \Psi (t) \rangle.
\end{equation}
Substituting Eq. \eqref{psiPsi} into the Eq. \eqref{TDSE}, the Hamiltonian operator $\hat{h}(t)$ can be written in terms of $\hat{H}(t)$ and $\hat{\eta}(t)$ in the form
\begin{equation}
\label{h0}
\hat{h}(t) = \hat{\eta}(t) \hat{H}(t) \hat{\eta}^{-1}(t) + i \partial_{t}\hat{\eta}(t)\hat{\eta}^{-1}(t),
\end{equation}
which is assumed to be Hermitian  in order to generate a unitary time-evolution. Therefore, from the hermiticity condition on Eq. \eqref{h0}, $\hat{h}^\dagger(t)=\hat{h}(t)$, we obtain the time-dependent quasi-hermiticity relation
\begin{equation}
\label{QHR}
\hat{H}^{\dagger}(t)\hat{\Theta}(t) - \hat{\Theta}(t)\hat{H}(t) = i\partial_{t}\hat{\Theta}(t),
\end{equation}
with $\hat{\Theta}(t)=\hat{\eta}^{\dagger}(t)\hat{\eta}(t)$ being the time-dependent metric operator \cite{fring2016a}. It ensures the time-dependent probability densities in the Hermitian and non-Hermitian systems to be related according to the equation
\begin{equation}
\langle \Psi(t)\vert \tilde{\Psi}(t) \rangle_{\hat{\Theta}(t)} = \langle \Psi(t) \vert \hat{\Theta}(t)\vert \tilde{\Psi}(t) \rangle = \langle \psi(t) \vert \tilde{\psi}(t) \rangle.
\end{equation}

Moreover, the Hermitian and non-Hermitian observables, respectively represented by $\hat{o}(t)$ and $\hat{O}(t)$, can be related through the similarity transformation
\begin{equation}
\label{obs}
\hat{o}(t) = \hat{\eta}(t)\hat{O}(t)\hat{\eta}^{-1}(t),
\end{equation}
in which the Dyson operator $\hat{\eta}(t)$ appears as the central element relating both operators. It also implies on the equality
\begin{equation}
\langle \Psi(t)\vert \hat{O}(t) \vert \tilde{\Psi}(t) \rangle_{\hat{\Theta}(t)} = \langle \psi(t) \vert \hat{o}(t)\vert \tilde{\psi}(t) \rangle,
\end{equation}
which shows that the equality of the expectation values in both non-Hermitian and Hermitian approaches. Unlike when the metric is time-dependent, in the case of time-independent metric operator, the Eqs. \eqref{h0} and \eqref{QHR} reduce to the forms
\begin{subequations}
\begin{align}
        \hat{h}(t) &= \hat{\eta}\hat{H}(t)\hat{\eta}^{-1},
        \\
        \hat{H}^{\dagger}(t) &= \hat{\Theta}\hat{H}(t)\hat{\Theta}^{-1},
\end{align}
\end{subequations}
which means that the Eq. \eqref{obs} holds, and therefore the Hamiltonian operator represents an observable in the quantum system.

In Ref. \cite{fring2017}, the authors discuss a remarkable issue arising from the case of time-dependent metric operator in the Hilbert space. The non-Hermitian Hamiltonian operator can not be associated with the observable correspondent to the energy operator, understood here only as the generator of time-evolution. This means that its eigenvalues are not necessarily real any time.
Nevertheless, a non-Hermitian observable related to the energy operator, $\tilde{H}(t)$, is defined from Eq. \eqref{obs} as
\begin{equation}\label{Hm}
        \tilde{H}(t) \equiv \hat{\eta}^{-1}(t)\hat{h}(t)\hat{\eta}(t) = \hat{H}(t) + i \hat{\eta}^{-1}(t)\partial_{t}\hat{\eta}(t),
\end{equation}
where $\tilde{H}(t)$ cannot be called "Hamiltonian", since it does not satisfy the Eq. \eqref{TDSE}.

\subsection{Non-Hermitian Hamiltonian and Dyson map}

For our main purpose we consider the non-Hermitian Hamiltonian operator in the following form
\begin{equation}
\label{HLie}
\hat{H}_{\mathfrak{s}}(t) = 2\omega(t)\hat{K}_{0} + 2\alpha(t)\hat{K}_{-}  + 2\beta(t)\hat{K}_{+},
\end{equation}
where the operator $\hat{K}_{i}$ is used to represent the $i$-th generators of $\mathfrak{su}(1,1)$ or $\mathfrak{su}(2)$ Lie algebras, which can be written in the unified form:
\begin{equation}
\label{LieAlgebras}
[\hat{K}_{0},\hat{K}_{\pm}] = \pm \hat{K}_{\pm}\,,
\quad
[\hat{K}_{+},\hat{K}_{-}] = 2\mathfrak{s}\hat{K}_{0} \, .
\end{equation}
Here the choice of the parameter $\mathfrak{s}=\pm1$ determines the correspondent $\mathfrak{su}(2)$ and $\mathfrak{su}(1,1)$, respectively. In the context of these Lie algebras, we now introduce the Hermitian time-dependent Dyson map defined in terms of the generators $\hat{K}_{i}$, and which has the form
\begin{subequations}
\begin{align}
\label{TDDM-Ks}
\hat{\eta}_{\mathfrak{s}}(t)
&=
\exp\big[2\epsilon_{\mathfrak{s}}\hat{K}_{0} + 2\mu_{\mathfrak{s}}\hat{K}_{-}  + 2\mu_{\mathfrak{s}}^{\ast}\hat{K}_{+}\big],
\\
\label{TDDM-K_GD}
&=
\exp\big[  \lambda_{\mathfrak{s}}\hat{K}_{+}\big]  \exp\big[  \ln\Lambda_{\mathfrak{s}}\hat{K}_{0}\big]  \exp\big[  \lambda_{\mathfrak{s}}^{\ast}\hat{K}_{-}\big].
\end{align}
\end{subequations}
Notice the parameters in both Eqs. \eqref{TDDM-Ks}-\eqref{TDDM-K_GD} are time dependent, and we suppres it in the notation for simplicity. The Eq. \eqref{TDDM-K_GD} comes from the Gauss decomposition \cite{ban1993} applied on Eq. \eqref{TDDM-Ks}, with $\lambda_{\mathfrak{s}}$ and $\Lambda_{\mathfrak{s}}$ conveniently written in the form
\begin{subequations}
        \label{TDDMlambdas}
        \begin{align}
        \label{TDDMlambda}
        \lambda_{\mathfrak{s}}
        &=
        \Phi_{\mathfrak{s}}e^{-i\varphi_{\mathfrak{s}}},
        \\
        \label{TDDMlambda0}
        \Lambda_{\mathfrak{s}}
        &=
        \frac{1-\tanh^{2}\Xi_{\mathfrak{s}}}{\left(
        1-\epsilon_{\mathfrak{s}}\tanh\Xi_{\mathfrak{s}}/\Xi_{\mathfrak{s}}\right)^{2}},
        \end{align}
\end{subequations}
with $\Lambda_{\mathfrak{s}} > 0$, and
\begin{equation}
\label{Phi}
\Phi_{\mathfrak{s}} = \frac{\epsilon_{\mathfrak{s}}\tanh{\Xi_{\mathfrak{s}}}/\Xi_{\mathfrak{s}}}{1-\epsilon_{\mathfrak{s}}\tanh{\Xi_{\mathfrak{s}}}/\Xi_{\mathfrak{s}}}|z_{\mathfrak{s}}|.
\end{equation}
In the Eq. \eqref{Phi} the parameter $z_{\mathfrak{s}}=2\mu_{\mathfrak{s}}/\epsilon_{\mathfrak{s}}=| z_{\mathfrak{s}}| e^{i\varphi_{\mathfrak{s}}}$ is known to be the free parameter of the map, where we are assuming the parameter $\epsilon_{\mathfrak{s}}$ as a positive real function and $\mu_{\mathfrak{s}}=|\mu_{\mathfrak{s}}|e^{i\varphi_{\mathfrak{s}}}$.
Also, we have the parameter $\Xi_{\mathfrak{s}}=\sqrt{\epsilon_{\mathfrak{s}}^{2}+4\mathfrak{s}|\mu_{\mathfrak{s}}|^{2}} = \epsilon_{\mathfrak{s}}\sqrt{1+\mathfrak{s}| z_{\mathfrak{s}}|^{2}}$ such that $\Xi_{\mathfrak{s}} \in \mathbb{R}$ by assumption.
Unlike the $\mathfrak{su}(2)$ case which does not imply any constraint to $|z_{\mathfrak{s}}|$, the assumption $\Xi_{\mathfrak{s}} \in \mathbb{R}$ implies that we need to consider the additional condition $|z_{\mathfrak{s}}|\leq1$ for the case of $\mathfrak{su}(1,1)$ Lie algebra.

The Eq. \eqref{Phi} allows to express the parameter $\epsilon_{\mathfrak{s}}$ in terms of $z_{\mathfrak{s}}$ and $\Phi_{\mathfrak{s}}$ in the form
\begin{align}
\label{epsilon}
\epsilon_{\mathfrak{s}} = \frac{1}{2\sqrt{1+\mathfrak{s}|z_{\mathfrak{s}}|^{2}}}\ln \frac{\big(1 + \sqrt{1+\mathfrak{s}|z_{\mathfrak{s}}|^{2}}\,\big)\Phi_{\mathfrak{s}} + |z_{\mathfrak{s}}|}{\big(1 - \sqrt{1+\mathfrak{s}|z_{\mathfrak{s}}|^{2}}\,\big)\Phi_{\mathfrak{s}} + |z_{\mathfrak{s}}|}.
\end{align}
Conveniently, we also introduce the real function
\begin{align}
\label{chi}
\chi_{\mathfrak{s}} &= -\mathfrak{s}\Phi_{\mathfrak{s}}^{2}-\Lambda_{\mathfrak{s}}
= -\frac{2\Phi_{\mathfrak{s}}}{|z_{\mathfrak{s}}|}-1,
\end{align}
which simplifies the notation. Also, it allows to obtain the modulus of the free-parameter $|z_{\mathfrak{s}}|$ in terms of the time-dependent parameters $\Phi_{\mathfrak{s}}$ and $\Lambda_{\mathfrak{s}}$,
\begin{equation}
\label{z}
|z_{\mathfrak{s}}| = \frac{2\Phi_{\mathfrak{s}}}{\Lambda_{\mathfrak{s}} + \mathfrak{s}\Phi_{\mathfrak{s}}^{2}-1}\,.
\end{equation}
We only have defined the Dyson map so far, and now we are able to build the Hermitian counterpart \eqref{h0}.

\subsection{Hermitian counterpart}

Since we are considering the structure of the non-Hermitian Hamiltonian operator in terms of the generators of the $\mathfrak{su}(2)$ and $\mathfrak{su}(1,1)$ Lie algebras, the operator correspondent to the Hermitian counterpart  $\hat{h}(t)$, obtained from the Eq. \eqref{h0}, have to be written in terms of same Lie algebra generators:
\begin{equation}
\label{h1}
\hat{h}_{\mathfrak{s}}(t) = 2\mathcal{W}_{\mathfrak{s}}(t)\hat{K}_{0} + 2\mathcal{U}_{\mathfrak{s}}(t)\hat{K}_{-} + 2\mathcal{V}_{\mathfrak{s}}(t)\hat{K}_{+}.
\end{equation}
To determine the explicty form of the coefficients in \eqref{h1}, we apply the following transformations
\begin{subequations}
\begin{align}
\hat{\eta}_{\mathfrak{s}}\hat{K}_{0} \hat{\eta}_{\mathfrak{s}}^{-1}
&=
\frac{\mathfrak{s}\Phi_{\mathfrak{s}}^{2}-\chi_{\mathfrak{s}}}{\Lambda_{\mathfrak{s}}}\hat{K}_{0} + \frac{\lambda_{\mathfrak{s}}^{\ast}}{\Lambda_{\mathfrak{s}}}\hat{K}_{-} + \frac{\lambda_{\mathfrak{s}} \chi_{\mathfrak{s}}}{\Lambda_{\mathfrak{s}}} \hat{K}_{+},
\\
\hat{\eta}_{\mathfrak{s}}\hat{K}_{-} \hat{\eta}_{\mathfrak{s}}^{-1}
&=
\frac{2\mathfrak{s}\lambda_{\mathfrak{s}}}{\Lambda_{\mathfrak{s}}}\hat{K}_{0} + \frac{1}{\Lambda_{\mathfrak{s}}}\hat{K}_{-} - \frac{\mathfrak{s}\lambda_{\mathfrak{s}}^{2}}{\Lambda_{\mathfrak{s}}}\hat{K}_{+},
\\
\hat{\eta}_{\mathfrak{s}}\hat{K}_{+} \hat{\eta}_{\mathfrak{s}}^{-1}
&=
\frac{2 \mathfrak{s}\lambda_{\mathfrak{s}}^{\ast}\chi_{\mathfrak{s}}}{\Lambda_{\mathfrak{s}}}\hat{K}_{0} - \frac{\mathfrak{s} \lambda_{\mathfrak{s}}^{\ast 2}}{\Lambda_{\mathfrak{s}}}\hat{K}_{-} + \frac{\chi_{\mathfrak{s}}^{2}}{\Lambda_{\mathfrak{s}}}\hat{K}_{+} ,
\end{align}
\end{subequations}
together the additional result correspondent to the term involving the time derivative on the Dyson map appearing in Eq. \eqref{Hm}:
\begin{align}
\partial_{t}\hat{\eta}_{\mathfrak{s}}\hat{\eta}_{\mathfrak{s}}^{-1}
=&
\frac{1}{\Lambda_{\mathfrak{s}}}
\big( \dot{\Lambda}_{\mathfrak{s}} + 2\mathfrak{s}\dot{\lambda}_{\mathfrak{s}}^{\ast}\lambda_{\mathfrak{s}} \big)\hat{K}_{0}
+
\frac{\dot{\lambda}_{\mathfrak{s}}^{\ast}}{\Lambda_{\mathfrak{s}}}\hat{K}_{-}
\nonumber\\
+&
\frac{1}{\Lambda_{\mathfrak{s}}}\big(\dot{\lambda}_{\mathfrak{s}}\Lambda_{\mathfrak{s}} - \dot{\Lambda}_{\mathfrak{s}}\lambda_{\mathfrak{s}} -\mathfrak{s} \dot{\lambda}_{\mathfrak{s}}^{\ast} \lambda_{\mathfrak{s}}^{2} \big) \hat{K}_{+}.
\end{align}
Here like everywhere else, the dot at the top of functions means time derivative. After simple algebraic manipulations, we obtain the time-dependent coefficients in the Eq. \eqref{h1},  which are written as
\begin{subequations}
\label{TDC-h}
\begin{align}
\mathcal{W}_{\mathfrak{s}}(t) &= \frac{1}{\Lambda_{\mathfrak{s}}}
\bigg[ \omega(\mathfrak{s}\Phi_{\mathfrak{s}}^{2} - \chi_{\mathfrak{s}} ) + 2\mathfrak{s}(\alpha\lambda_{\mathfrak{s}} + \beta\lambda_{\mathfrak{s}}^{\ast}\chi_{\mathfrak{s}})
\nonumber\\
&+ \frac{i}{2}(\dot{\Lambda}_{\mathfrak{s}} + 2\mathfrak{s}\dot{\lambda}_{\mathfrak{s}}^{\ast}\lambda_{\mathfrak{s}}) \bigg],
\\
\mathcal{U}_{\mathfrak{s}}(t) &= \frac{1}{\Lambda_{\mathfrak{s}}} \bigg[  \omega\lambda_{\mathfrak{s}}^{\ast} + \alpha - \mathfrak{s}\beta \lambda_{\mathfrak{s}}^{\ast 2} + i\frac{\dot{\lambda}_{\mathfrak{s}}^{\ast}}{2} \bigg],
\\
\mathcal{V}_{\mathfrak{s}}(t) &= \frac{1}{\Lambda_{\mathfrak{s}}}
\bigg[\omega\lambda_{\mathfrak{s}}\chi_{\mathfrak{s}} - \mathfrak{s}\alpha\lambda_{\mathfrak{s}}^{2} + \beta\chi_{\mathfrak{s}}^{2}
\nonumber\\
&+ \frac{i}{2}(\dot{\lambda}_{\mathfrak{s}}\Lambda_{\mathfrak{s}} - \dot{\Lambda}_{\mathfrak{s}}\lambda_{\mathfrak{s}} - \mathfrak{s} \dot{\lambda}_{\mathfrak{s}}^{\ast} \lambda_{\mathfrak{s}}^{2}) \bigg].
\end{align}
\end{subequations}
Notice the obtained Hamiltonian operator from transformation Eq. \eqref{h0} is not an Hermitian in general, due to the algebraic structure of the Dyson mapping. We have to impose the hermiticity condition on $\hat{h}_{\mathfrak{s}}(t)$, reads as $\hat{h}_{\mathfrak{s}}(t)=\hat{h}_{\mathfrak{s}}^{\dagger}(t)$, which leads to the following relations
\begin{equation}
\label{constraint}
\mathcal{W}_{\mathfrak{s}}(t)=\mathcal{W}_{\mathfrak{s}}^{\ast}(t), \qquad \mathcal{V}_{\mathfrak{s}}(t)=\mathcal{U}_{\mathfrak{s}}^{\ast}(t).
\end{equation}
Thus, the hermiticity condition on $\hat{h}_{\mathfrak{s}}(t)$ implies to the Hermitian counterpart of the operator \eqref{HLie} the restrictive form
\begin{equation}
\label{h}
\hat{h}_{\mathfrak{s}}(t) = 2\mathcal{W}_{\mathfrak{s}}(t)\hat{K}_{0} + 2\mathcal{U}_{\mathfrak{s}}(t)\hat{K}_{-} + 2\mathcal{U}_{\mathfrak{s}}^{\ast}(t)\hat{K}_{+},
\end{equation}
wherein the time-dependent coefficients $\mathcal{W}_{\mathfrak{s}}(t)$ and  $\mathcal{U}_{\mathfrak{s}}(t)$: 
\begin{subequations}
\label{WU}
\begin{align}
\label{W}
\mathcal{W}_{\mathfrak{s}}(t)
&=
\omega_{\text{R}}
- \frac{2\mathfrak{s}\Phi_{\mathfrak{s}}}{\chi_{\mathfrak{s}} - 1}
\nonumber\\
&\times\left[|\alpha|\cos{(\varphi_{\mathfrak{s}}-\varphi_{\alpha})}
- |\beta|\cos{(\varphi_{\mathfrak{s}}+\varphi_{\beta})}\right],
\\
\label{U}
\mathcal{U}_{\mathfrak{s}}(t)
&=
\frac{1}{1 - \chi_{\mathfrak{s}}}
\left[
\alpha - \chi_{\mathfrak{s}}\beta^{\ast} + ie^{i\varphi_{\mathfrak{s}}}\Phi_{\mathfrak{s}}\omega_{\text{I}}
\right].
\end{align}
\end{subequations}
Here $\alpha = |\alpha|e^{i\varphi_{\alpha}}$, $\beta = |\beta|e^{i\varphi_{\beta}}$ and $\omega = \omega_{\text{R}} + i \omega_{\text{I}}$
with $\omega_{\text{R}}$ and $\omega_{\text{I}}$ the real and imaginary part of $\omega$, respectively. The Dyson map parameters are now constrained to the following set of coupled nonlinear differential equations:
\begin{widetext}
\begin{subequations}
\label{DysonMapEqs}
\begin{align}
\dot{\Phi}_{\mathfrak{s}}
&=
\frac{2}{\chi_{\mathfrak{s}} - 1}
\bigg\{
[\Phi_{\mathfrak{s}}\omega_{\text{I}} - |\alpha|\sin{(\varphi_{\mathfrak{s}}-\varphi_{\alpha})}](1 + \mathfrak{s}\Phi_{\mathfrak{s}}^{2})
+ |\beta|\sin{(\varphi_{\mathfrak{s}} + \varphi_{\beta})}[\mathfrak{s}(2\chi_{\mathfrak{s}} - 1)\Phi_{\mathfrak{s}}^{2} + \chi_{\mathfrak{s}}^{2}]
\bigg\},
\\
\dot{\varphi}_{\mathfrak{s}}
&=
2\omega_{\text{R}}
-
\frac{2}{(\chi_{\mathfrak{s}}-1) \Phi_{\mathfrak{s}}}
\bigg\{
|\alpha|\cos{(\varphi_{\mathfrak{s}}-\varphi_{\alpha})}(1 + \mathfrak{s}\Phi_{\mathfrak{s}}^{2})
-
|\beta|\cos{(\varphi_{\mathfrak{s}}+\varphi_{\beta})}(\mathfrak{s}\Phi_{\mathfrak{s}}^{2} + \chi_{\mathfrak{s}}^{2} )
\bigg\},
\\
\dot{\Lambda}_{\mathfrak{s}}&=
2\Lambda_{\mathfrak{s}}\left\{ \left( \frac{2\mathfrak{s}\Phi_{\mathfrak{s}}^{2}}{\chi_{\mathfrak{s}}-1}-1\right) \omega_{\text{I}}
-
\frac{2\mathfrak{s}\Phi_{\mathfrak{s}}}{\chi_{\mathfrak{s}}-1}\left[  |\alpha| \sin\left(  \varphi_{\mathfrak{s}}-\varphi_{\alpha }\right)  -|\beta| \left(  2\chi_{\mathfrak{s}}-1\right) \sin\left(  \varphi_{\mathfrak{s}} + \varphi_{\beta}\right)  \right]  \right\},
\end{align}
\end{subequations}
\end{widetext}
which arise due to the hermiticity conditions expressed in Eq. \eqref{constraint}. Otherwise, we reinforce that the Hamiltonian \eqref{h} is Hermitian for each Dyson map parameter which satisfies the Eq. \eqref{DysonMapEqs}. Also, as in Refs. \cite{deponte2019,deponte2021E,dourado2021},  we refers to $z_{\mathfrak{s}}$ as being the only free parameter determining the Dyson map, with $\Phi_{\mathfrak{s}}$, $\varphi_{\mathfrak{s}}$ and $\Lambda_{\mathfrak{s}}$ coming from the set of coupled equations \eqref{DysonMapEqs}, and $\epsilon_{\mathfrak{s}}$ coming from Eq. \eqref{epsilon}. This implies that a given pair ($|z_{\mathfrak{s}}|$, $\Phi_{\mathfrak{s}}$) must be further corroborated by a real and positive $\epsilon_{\mathfrak{s}}$.

The results described up to Eq. \eqref{TDC-h} were obtained in Refs. \cite{maamache2017,koussa2018} for which the authors just considered the simplest case $\mathcal{V}_{\mathfrak{s}}(t)=\mathcal{U}_{\mathfrak{s}}(t)=0$. On the other hand, we have imposed more general conditions \eqref{constraint} to achieve the hermiticity of $\hat{h}_{\mathfrak{s}}(t)$. Furthermore, notice that for the case of the $\mathfrak{su}(1,1)$ Lie algebra, our results reproduce exactly the same obtained in \cite{fring2016b,deponte2019,deponte2021E}, where the authors considered the one-mode realization of the Lie algebra, what reinforces the fact that the Hermitian counterpart is independent of the Lie algebra realization, such as in the time-independent case treated in Ref. \cite{quesne2007}.

\subsection{Time-evolution}
The first step in studying the dynamic behavior of the system described by the non-Hermitian Hamiltonian operator \eqref{HLie} was given by obtaining the associated  Hermitian Hamiltonian operator  \eqref{h} by means of the time-dependent Dyson mapping.  From now on, our discussion concerns the procedure for solving the Schrödinger equation \eqref{TDSE-h}. For this purpose, we follow a similar strategy adopted in Ref. \cite{mizrahi1994}, whereby we have to consider the transformation
\begin{equation}
\label{OriginalSolution}
\vert \psi_{\mathfrak{s}}(t) \rangle = \hat{S}_{\mathfrak{s}}(t) \vert \tilde{\psi}_{\mathfrak{s}}(t) \rangle,
\end{equation}
with the unitary operator $\hat{S}_{\mathfrak{s}}(t)$ defined as follows
\begin{equation}
\label{Sop}
\hat{S}_{\mathfrak{s}}(t) =\exp\big[\xi_{\mathfrak{s}}(t)\hat{K}_{+} - \xi_{\mathfrak{s}}^{\ast}(t)\hat{K}_{-}\big].
\end{equation}
In Eq. \eqref{Sop} the time-dependent parameter $\xi_{\mathfrak{s}}(t) = r_{\mathfrak{s}}(t)e^{i\phi_{\mathfrak{s}}(t)}$ with $r_{\mathfrak{s}}(t),\phi_{\mathfrak{s}}(t)$ real functions and $r_{\mathfrak{s}}(t)\geq0$.
By substituting the Eq. \eqref{OriginalSolution} into Eq. \eqref{TDSE-h}, we obtain the Schrödinger equation for the vector state $ \vert \tilde{\psi}_{\mathfrak{s}}(t) \rangle$
\begin{equation}
\label{SE_Htil}
i\partial_{t} \vert \tilde{\psi}_{\mathfrak{s}}(t) \rangle = \hat{\mathcal{H}}_{\mathfrak{s}}(t) \vert \tilde{\psi}_{\mathfrak{s}}(t) \rangle,
\end{equation}
for which  the transformed Hamiltonian reads
\begin{equation}
\label{Htil}
\hat{\mathcal{H}}_{\mathfrak{s}}(t) = \hat{S}_{\mathfrak{s}}^{-1} \hat{h}_{\mathfrak{s}} \hat{S}_{\mathfrak{s}} + i\partial_{t}\hat{S}_{\mathfrak{s}}^{-1}\hat{S}_{\mathfrak{s}}.
\end{equation}
After straightforward calculations, the transformed Hamiltonian $\hat{\mathcal{H}}_{\mathfrak{s}}$ reduces to
\begin{equation}
\hat{\mathcal{H}}_{\mathfrak{s}}(t) = 2\Omega_{\mathfrak{s}}(t)\hat{K}_{0},
\end{equation}
with
\begin{align}
\label{R-parameter}
\Omega_{\mathfrak{s}} &=
\mathcal{W}_{\mathfrak{s}}
-
2\sqrt{-\mathfrak{s}}|\mathcal{U}_{\mathfrak{s}}|
\tanh{(\sqrt{-\mathfrak{s}}r_{\mathfrak{s}})}
\cos{(\phi_{\mathfrak{s}} + \varphi_{\mathcal{U}_{\mathfrak{s}}})},
\end{align}
and  the parameters $r_{\mathfrak{s}}$ and $\phi_{\mathfrak{s}}$ having to satisfy the set of coupled differential equations
\begin{subequations}
        \label{SqueezingParametersDE}
        \begin{align}
        \label{r0}
        \dot{r}_{\mathfrak{s}} &= -2|\mathcal{U}_{\mathfrak{s}}|\sin{(\phi_{\mathfrak{s}}+\varphi_{\mathcal{U}_{\mathfrak{s}}})},
        \\
        \label{phi0}
        \dot{\phi}_{\mathfrak{s}} &=-2\mathcal{W}_{\mathfrak{s}}
        \nonumber\\ &-4\sqrt{-\mathfrak{s}}|\mathcal{U}_{\mathfrak{s}}|\coth{(2\sqrt{-\mathfrak{s}}r_{\mathfrak{s}})}\cos{(\phi_{\mathfrak{s}}+\varphi_{\mathcal{U}_{\mathfrak{s}}})},
        \end{align}
\end{subequations}
where we have adopted the polar form $\mathcal{U}_{\mathfrak{s}}=|\mathcal{U}_{\mathfrak{s}}|e^{i\varphi_{\mathcal{U}_{\mathfrak{s}}}}$. Therefore, the formal solution of Eq. \eqref{SE_Htil} becomes
\begin{equation}
\vert \tilde{\psi}_{\mathfrak{s}}(t) \rangle = \hat{R}_{\mathfrak{s}}(t)\vert \tilde{\psi}_{\mathfrak{s}}(0) \rangle,
\end{equation}
in which $\hat{R}_{\mathfrak{s}}(t)$ is defined as
\begin{equation}
                \hat{R}_{\mathfrak{s}}(t) =\exp\big[-2i\tilde{\Omega}_{\mathfrak{s}}(t)\hat{K}_{0}\big],
\end{equation}
with the time-dependent function $\tilde{\Omega}_{\mathfrak{s}}(t)=\int_{0}^{t}d\tau\Omega_{\mathfrak{s}}(\tau)$. Finally, the formal solution to Eq. \eqref{TDSE-h} reads
\begin{equation}
\label{state-t}
\vert \psi_{\mathfrak{s}}(t) \rangle = \hat{u}_{\mathfrak{s}}(t)\vert \psi_{\mathfrak{s}}(0) \rangle,
\end{equation}
where the unitary time-evolution operator $\hat{u}_{\mathfrak{s}}(t)$ is given by
\begin{align}
\label{evol-u}
\hat{u}_{\mathfrak{s}}(t) =
\hat{S}_{\mathfrak{s}}(t)\hat{R}_{\mathfrak{s}}(t)\hat{S}_{\mathfrak{s}}^{\dagger}(0).
\end{align}

Furthermore, the solution of Schrödinger equation \eqref{TDSE-H} becomes
\begin{align}
\vert \Psi_{\mathfrak{s}} (t) \rangle &= \hat{\eta}_{\mathfrak{s}}^{-1}(t) \hat{u}_{\mathfrak{s}}(t)\vert \psi_{\mathfrak{s}} (0) \rangle
\nonumber\\
&=  \hat{\eta}_{\mathfrak{s}}^{-1}(t) \hat{u}_{\mathfrak{s}}(t)\hat{\eta}_{\mathfrak{s}}(0)\vert \Psi_{\mathfrak{s}} (0) \rangle ,
\end{align}
from where we indentiy the time evolution operator $\hat{U}_{\mathfrak{s}}(t)$ as
\begin{equation}
\label{evol-U}
    \hat{U}_{\mathfrak{s}}(t) = \hat{\eta}_{\mathfrak{s}}^{-1}(t) \hat{u}_{\mathfrak{s}}(t)\hat{\eta}_{\mathfrak{s}}(0)\,.
\end{equation}
After discussing the formal apparatus establishing the connection between non-Hermitian aspects of the system dynamics with its Hermitian counterpart, we are able to analyze any cases associated with the Hamiltonian operator \eqref{HLie} without considering the algebra realization.

\section{The $\hat{K}_{0}$ case}
\label{sec:K0}
The simplest model associated with the non-Hermitian Hamiltonian \eqref{HLie} is obtained assuming the time-dependent coefficients $\alpha(t)=\beta(t)=0$, such that the Hamiltonian operator has the form
\begin{equation}
\label{k0nh}
\hat{H}_{\mathfrak{s}}(t) = 2\omega(t)\hat{K}_{0}.
\end{equation}
The correspondent Hermitian counterpart \eqref{h} becomes
\begin{equation}
\label{k0h}
\hat{h}_{\mathfrak{s}}(t) = 2\omega_{\text{R}}\hat{K}_{0} + 2i\frac{\Phi_{\mathfrak{s}}\omega_{\text{I}}}{1 - \chi_{\mathfrak{s}}  }
[e^{i\varphi_{\mathfrak{s}}}\hat{K}_{-} - e^{-i\varphi_{\mathfrak{s}}}\hat{K}_{+}],
\end{equation}
wherein the time-dependent coefficients from Eq. \eqref{WU} read $\mathcal{W}_{\mathfrak{s}}(t)=\omega_{\text{R}}$ and $\mathcal{U}_{\mathfrak{s}}(t)=ie^{i\varphi_{\mathfrak{s}}}\Phi_{\mathfrak{s}}\omega_{\text{I}}/(1 - \chi_{\mathfrak{s}} )$. The time-dependent Dyson map parameters given by Eq. \eqref{DysonMapEqs} reduce to the following forms
\begin{subequations}
        \label{EXDysonMapEqs}
        \begin{align}
        \label{PhiDE}
        \dot{\Phi}_{\mathfrak{s}}
        &=
        \frac{2\Phi_{\mathfrak{s}}(1 + \mathfrak{s}\Phi_{\mathfrak{s}}^{2})}{\chi_{\mathfrak{s}} - 1}
        \omega_{\text{I}},
        \\
        \dot{\varphi}_{\mathfrak{s}}
        &=
        2\omega_{\text{R}},
        \\
        \dot{\Lambda}_{\mathfrak{s}}&=
        2\Lambda_{\mathfrak{s}}
        \left( \frac{2\mathfrak{s}\Phi_{\mathfrak{s}}^{2}}{\chi_{\mathfrak{s}}-1}-1\right) \omega_{\text{I}}.
        \end{align}
\end{subequations}
The solutions of these set of differential equations are given by
\begin{subequations}
\label{DysonMapParametersSol}
        \begin{align}
        \Phi_{\mathfrak{s}}(t)
        &= \Phi_{\mathfrak{s}}(0)\frac{ \Lambda_{\mathfrak{s}}(0) + \mathfrak{s}\Phi_{\mathfrak{s}}^{2}(0) + 1}{\Lambda_{\mathfrak{s}}(0)
                + [\mathfrak{s}\Phi_{\mathfrak{s}}^{2}(0) + 1]e^{2\int_{0}^{t}d\tau\omega_{\text{I}}(\tau)}},
        \label{PhiSol}
        \\
        \varphi_{\mathfrak{s}}(t)
        &=
        \varphi_{\mathfrak{s}}(0) +
        2\int_{0}^{t}d\tau\omega_{\text{R}}(\tau),
        \\
        \Lambda_{\mathfrak{s}}(t) &= \Lambda_{\mathfrak{s}}(0)\frac{\mathfrak{s}\Phi_{\mathfrak{s}}^{2}(t) + 1}{\mathfrak{s}\Phi_{\mathfrak{s}}^{2}(0) + 1}
        e^{-2\int_{0}^{t}d\tau\omega_{\text{I}}(\tau)},
        \label{LambdaSol}
        \end{align}
\end{subequations}
whereas the modulus of the free parameter $z_{\mathfrak{s}}$ becomes completely defined by the Eq. \eqref{z}.

In order to obtain the time-evolution of the quantum system, we have to solve the set of differential equations expressed in Eq. \eqref{SqueezingParametersDE}. For this purpose, notice that
\[
\mathcal{U}_{\mathfrak{s}}(t)
=|\mathcal{U}_{\mathfrak{s}}|e^{i\varphi_{\mathcal{U}_{\mathfrak{s}}}}
=-i\frac{\Phi_{\mathfrak{s}}}{\chi_{\mathfrak{s}} - 1}\omega_{\text{I}}e^{i\varphi_{\mathfrak{s}}},
\]
leads to the identities
\begin{subequations}\label{Sd}
\begin{align}
|\mathcal{U}_{\mathfrak{s}}|\cos\varphi_{\mathcal{U}_{\mathfrak{s}}} &= \frac{\Phi_{\mathfrak{s}}\omega_{\text{I}}}{\chi_{\mathfrak{s}} - 1}\sin\varphi_{\mathfrak{s}},
\\
|\mathcal{U}_{\mathfrak{s}}|\sin\varphi_{\mathcal{U}_{\mathfrak{s}}} &= -\frac{\Phi_{\mathfrak{s}}\omega_{\text{I}}}{\chi_{\mathfrak{s}} - 1}\cos\varphi_{\mathfrak{s}},
\end{align}
\end{subequations}
which allows us to rewrite Eq. \eqref{SqueezingParametersDE} as
\begin{subequations}
        \label{SqueezingParametersDE2}
        \begin{align}
        \label{r02}
        \dot{r}_{\mathfrak{s}} &= \frac{2\Phi_{\mathfrak{s}}}{\chi_{\mathfrak{s}} - 1}\omega_{\text{I}}\cos{(\varphi_{\mathfrak{s}}+\phi_{\mathfrak{s}})},
        \\
        \label{phi02}
        \dot{\phi}_{\mathfrak{s}} &=-2\omega_{\text{R}}
        \nonumber\\ &-\frac{4\sqrt{-\mathfrak{s}}\Phi_{\mathfrak{s}}\omega_{\text{I}}}{\chi_{\mathfrak{s}} - 1}\coth{(2\sqrt{-\mathfrak{s}}r_{\mathfrak{s}})}\sin{(\varphi_{\mathfrak{s}}+\phi_{\mathfrak{s}})}.
        \end{align}
\end{subequations}
A considerable simplification is brought for Eqs. \eqref{Sd} if we assume that
\begin{equation}
\varphi_{\mathfrak{s}} = l\pi - \phi_{\mathfrak{s}} ,
\end{equation}
which implies in the maximum rate of change in time of the function $r_{\mathfrak{s}}$ for a given  $\mathcal{U}_{\mathfrak{s}}$.  Furthermore, from the Eq. \eqref{PhiDE}, we obtain the following equality
\[
\frac{2\Phi_{\mathfrak{s}}}{\chi_{\mathfrak{s}}-1 }\omega_{\text{I}}
=\frac{\dot{\Phi}_{\mathfrak{s}}}{1+\mathfrak{s}\Phi_{\mathfrak{s}}^{2}},
\]
which leads to the simpler set of differential equation
\begin{subequations}
        \label{SqueezingParametersDE3}
        \begin{align}
        \label{r03}
        \dot{r}_{\mathfrak{s}} &= (-1)^{l}\frac{\dot{\Phi}_{\mathfrak{s}}}{1+\mathfrak{s}\Phi_{\mathfrak{s}}^{2}},
        \\
        \label{phi03}
        \dot{\phi}_{\mathfrak{s}} &=-2\omega_{\text{R}}.
        \end{align}
\end{subequations}
We note that both equations in Eq. \eqref{SqueezingParametersDE3} become uncoupled, and they can be directly integrated obtaining the solutions in the form
\begin{subequations}
        \label{SqueezingParametersSol}
        \begin{align}
        \label{rSol}
        r_{\mathfrak{s}}(t) &= r_{\mathfrak{s}}(0) + \frac{(-1)^{l}}{2\sqrt{-\mathfrak{s}}}
        \ln{\frac{\varpi_{\mathfrak{s}}(t)}{\varpi_{\mathfrak{s}}(0)}},
        \\
        \label{phiSol}
        \phi_{\mathfrak{s}}(t) &= l\pi -\varphi_{\mathfrak{s}}(0) - 2\int_{0}^{t}d\tau\omega_{\text{R}}(\tau),
        \end{align}
\end{subequations}
with the function $\varpi_{\mathfrak{s}}(t)$ reading
\begin{equation}
\label{varpi}
\varpi_{\mathfrak{s}}(t) = \frac{1+\sqrt{-\mathfrak{s}}\Phi_{\mathfrak{s}}(t)}{1-\sqrt{-\mathfrak{s}}\Phi_{\mathfrak{s}}(t)}.
\end{equation}
The integer $l$ can be chosen in order to make $r_{\mathfrak{s}}>0$ everytime. Furthermore, to determine the unitary time-evolution operator in Eq. \eqref{evol-u}, we still have to calculate the parameter $\Omega_{\mathfrak{s}}(t)$ given in Eq. \eqref{R-parameter}, which yields
\begin{equation}
\Omega_{\mathfrak{s}}(t) = \omega_{\text{R}},
\end{equation}
under the previous assumptions. Although $r_{\mathfrak{s}}(t)$ seems to be a complex function when $\mathfrak{s}=1$, straightforward calculations show that the complex logarithm may be rewritten in terms of the arc-tangent multi-valued real function. 
These solutions are similar to those obtained in \cite{dourado2021} by
considering the single-mode realization of $\mathfrak{su}(1,1)$, and
correspond to the well-known time-dependent Swanson oscillator with a
pure imaginary time-dependent frequency ($\omega_{\text{R}}=0$). 
The authors devise this kind of system by taking into account an oscillator with a strongly parametric quadratic pumping.   

As mentioned before, we reinforce that the obtained solutions up to this point in our discussion are independent of the Lie algebra realizations. Focusing our interest on the entanglement occurring in the non-Hermitian scenario, in what follows, we consider the two-mode bosonic realizations of Lie algebras, and make a qualitative discussion on how to generalize our investigation to the multimode non-interacting case.

\section{Two uncoupled modes and entanglement}
\label{sec:entanglement}

The $\mathfrak{su}(1,1)$ and $\mathfrak{su}(2)$ Lie algebras have immediate relevance on issues concerning the nonclassical properties of light in the context of quantum optics \cite{gerry1991,karimi:14,hach:18}. For instance, Lie-group-theoretical approach is applied to analyze $SU(1,1)$ and $SU(2)$ interferometers in Ref. \cite{yurke1986}. Also, the bosonic realizations of $\mathfrak{su}(1,1)$ are applied to describe the (non)degenerate parametric amplifier \cite{heidmann1987,zhang2017, abdalla2007,mohamed2019}, while beam splitters \cite{campos1989,paris1999,kim2002} are described by means of the $\mathfrak{su}(2)$ Lie algebra.  Important results concerning the multimode bosonic realizations of Lie algebras are obtained which are relevant for generalized coherent states 
\cite{Lo1993,sollie1993,Lo1994,Lo1995}.
In what follows, we restrict our analysis to the usual two-mode bosonic realizations of the three elements of $\mathfrak{su}(1,1)$ and $\mathfrak{su}(2)$ as defined, for example, in Ref. \cite{wodkiewicz1985}.

As a measure of entanglement, the linear entropy is determined from
\begin{equation}
\label{linearentropy}
\mathcal{S}_{\mathfrak{s}}(t) = 1 - \mathrm{Tr}[\hat{\rho}_{\mathfrak{s}}^{(1)}(t)]^{2},
\end{equation}
with $\hat{\rho}_{\mathfrak{s}}^{(1)}(t)$ being the reduced density matrix for the first mode, which is obtained by the sum over all the degree of freedom of second mode, \textit{i.e.},  $\hat{\rho}_{\mathfrak{s}}^{(1)}(t) = \mathrm{Tr}_{2} \vert \psi_{\mathfrak{s}}(t)\rangle \langle \psi_{\mathfrak{s}} (t)\vert$. In fact, it corresponds to an approximation of the well-known von-Neumann entropy \cite{horodecki2009}.

In addition, hereafter, we adopt the following notation: when we explicit the $\mathfrak{s}$ value, rather than use the subscript $\mathfrak{s}=\pm1$ in functions and operators we index them by the minus or plus ($-$ or $+$) in according to the sign of $\mathfrak{s}$.

\subsection{$\mathfrak{su}(1,1)$ entanglement}
We have the elements of $\mathfrak{su}(1,1)$ related with the bosonic operators $\hat{a}_{i}$ and $\hat{a}_{j}^{\dagger}$ which satisfy the Weyl-Heisenberg algebra \cite{thangavelu1998},
\begin{equation}
[\hat{a}_{i},\hat{a}_{j}]=[\hat{a}_{i}^{\dagger},\hat{a}_{j}^{\dagger}]=0, \quad  [\hat{a}_{i},\hat{a}_{j}^{\dagger}]=\delta_{ij}.
\end{equation}
The relations are written as \cite{wodkiewicz1985,gerry1991}
\begin{subequations}
\label{2msu11}
\begin{align}
\hat{K}_{0} &= \frac{1}{2}(\hat{a}_{1}^{\dagger}\hat{a}_{1} + \hat{a}_{2}^{\dagger}\hat{a}_{2} + 1),
\qquad
\\
\hat{K}_{+}^{\dagger}&=\hat{K}_{-} = \hat{a}_{1}\hat{a}_{2},
\end{align}
\end{subequations}
and it is just the well-known realization commonly applied on description of the two-mode squeezing of light fields \cite{karimi:14}.

For our purpose, we consider the initial separable two-mode vacua state $\vert \psi_{-}(0) \rangle =\vert 0,0 \rangle$ for which the condition $r_{-}(0)=0$ is true, and for which the Eq. \eqref{state-t} provides
\begin{equation}
\vert \psi_{-}(t) \rangle = \frac{e^{-i\tilde{\Omega}_{-}}}{\cosh{r_{-}}}\sum_{n=0}^{\infty}e^{in\phi_{-}}\tanh^{n}r_{-}\vert n,n \rangle ,
\end{equation}
and the reduced density matrix becomes
\begin{equation}
\hat{\rho}_{-}^{(1)}(t) =
\sum_{n=0}^{\infty}\frac{\tanh^{2n} r_{-}}{\cosh^{2}r_{-}}\vert n \rangle \langle n \vert ,
\end{equation}
and the correpondent linear entropy \eqref{linearentropy} is given by
\begin{equation}
\label{Sl}
\mathcal{S}_{-}(t) = 1 - \sech[2r_{-}(t)].
\end{equation}
From the Eq. \eqref{Sl} we see that the entanglement is zero when $r_{-}(t)=0$, which is true for the initial state $\vert \psi_{-}(0) \rangle =\vert 0,0 \rangle$. The enhancement of the entanglement is always verified whenever the value of $r_{-}(t)$ increases in time, and we note that the maximum entanglement occurs with infinite squeezing, what seems to be possible as recently demonstrated by Dourado and co-workers \cite{dourado2021}. According to them, an infinite squeezing into a finite time interval is possible, for the bosonic one-mode realization of $\mathfrak{su}(1,1)$, by considering a time-dependent pure imaginary frequency generated by a strong parametric pumping. Since the results represented in Eqs. \eqref{DysonMapParametersSol} and \eqref{SqueezingParametersSol} are independent of the realization, our solutions correspond to their ones by applying the change $\Phi_{-}(t)\rightarrow-\Phi_{-}(t)$ for all the  results, and taking $l=1$ at Eq. \eqref{rSol}.

Let us consider the imaginary part of complex frequency being linear in time such that
\begin{equation}
\label{omega}
\omega_{\text{I}} (t) = \gamma^{2} t.
\end{equation}
In what follows, we examine the validity of time-dependent Dyson map parameters expressed in \eqref{DysonMapParametersSol} under the above imaginary part of frequency by changing $\Phi_{-}\rightarrow-\Phi_{-}$ in all previous results. Actually, this transformation is just convenient for recovering the results obtained in Ref. \cite{dourado2021}. Further, the authors demonstrated that there is a region where the modulus of the free parameter exceeds the unity. In this region, the hermitization process fails since we assume initially the condition $|z_{-}(t)|\leqslant1$.

\begin{figure}[h!]
        \includegraphics[width=\linewidth]{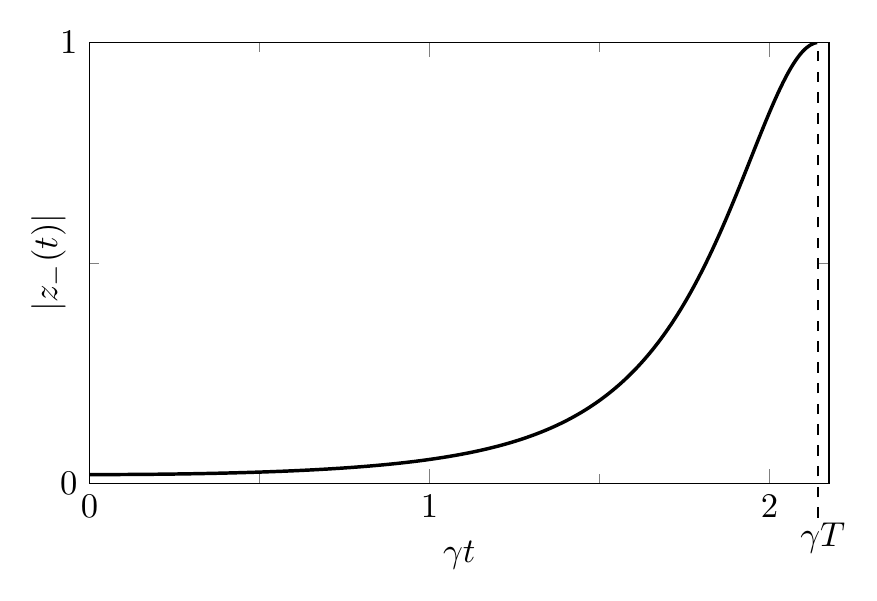}	\caption{\label{fig:z-} The modulus of free Dyson map parameter $|z_{-}(t)|$ in the dimensionless time $\gamma t$ with $\gamma=1/2\, \mathtt{s}^{-1}$ for the initial values $\Phi_{-}(0) = 10^{2}$ and $\Lambda_{-}(0) = 10^{-2}$. We evaluate the time-evolution until the dimensionless time $\gamma T \approx 2.15 $.}
\end{figure}

\begin{figure}[ht!]
        \includegraphics[width=\linewidth]{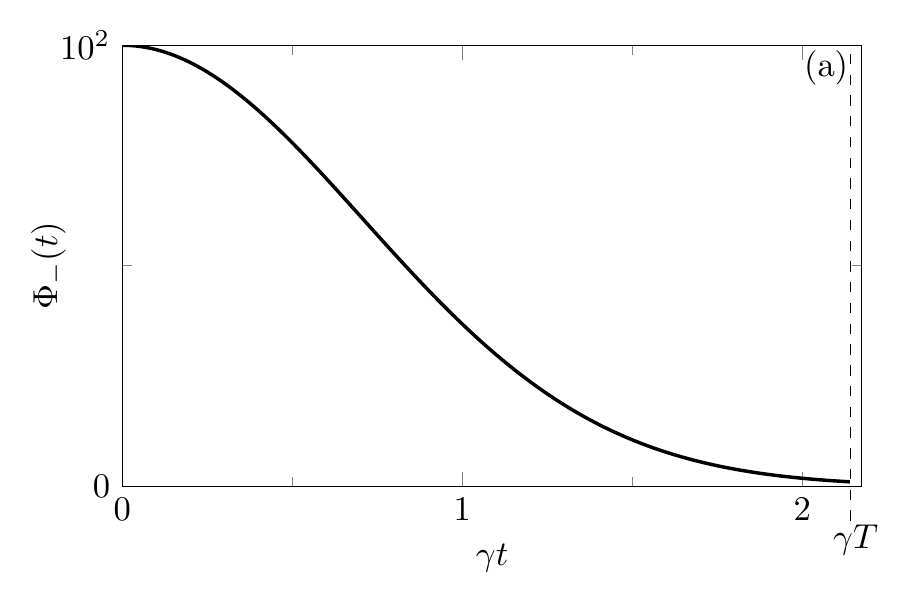}
        \includegraphics[width=\linewidth]{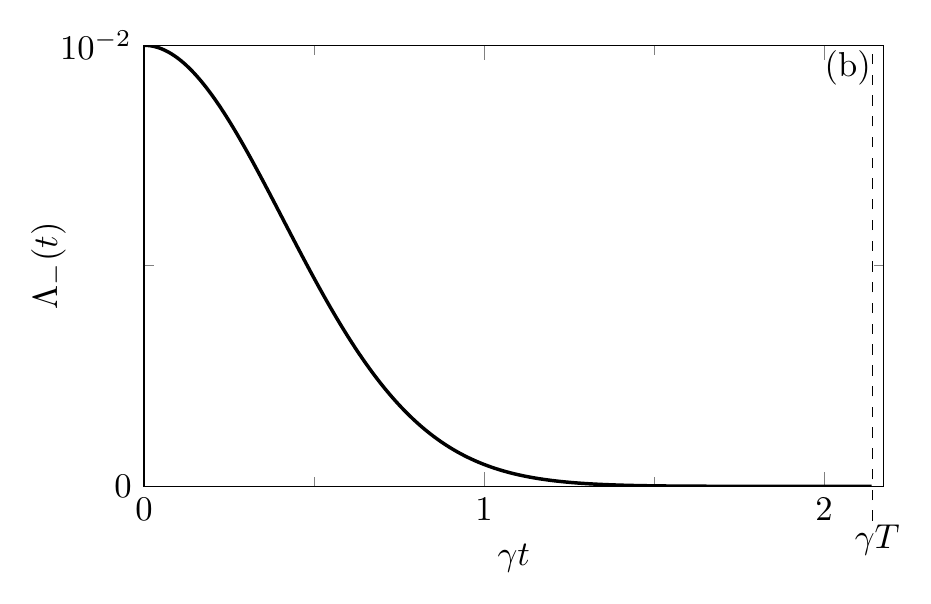}
\caption{\label{fig:PhiLambda-} The Dyson map parameters (a) $\Phi_{-}(t)$ and (b) $\Lambda_{-}(t)$  against the dimensionless time $\gamma t$ with $\gamma=1/2\, \mathtt{s}^{-1}$ for the initial values $\Phi_{-}(0) = 10^{2}$ and $\Lambda_{-}(0) = 10^{-2}$. We evaluate the time-evolution until the dimensionless time $\gamma T \approx 2.15 $.}
\end{figure}

In order to perform a more rigorous analysis, we verify that the modulus of free parameter $|z(t)|$ becomes equal to the unity at two different times, namely, $T_{\pm}$  given by
\begin{equation}
\label{int2}
T_{\pm}
=
\frac{1}{\gamma}\sqrt{\ln\left[\Phi_{-}(0) - \frac{\Lambda_{-}(0)}{\Phi_{-}(0)\pm 1}\right]},
\end{equation}
and then we calculate the time $T$ at which the maximum value of $|z_{-}(t)|$ occurs by solving the equation
\begin{equation*}
\frac{d|z_{-}(t)|}{dt}\bigg\vert_{t=T}=0.
\end{equation*}
It provides
\begin{equation}
T =
\frac{\sqrt{2}}{2}\sqrt{T_{-}^{2}+T_{+}^{2}},
\end{equation}
and at this time, the free parameter can be read as
\begin{equation}
\label{zmax}
|z_{-}(T)| =
\frac{1-\frac{\Lambda_{-}(0)}{\Phi_{-}^{2}(0)}-\frac{1}{\Phi_{-}^{2}(0)}}
{\sqrt{\Big[\left(1-\frac{\Lambda_{-}(0)}{\Phi_{-}^{2}(0)}\right)^{2}-\frac{1}{\Phi_{-}^{2}(0)}\Big]\left[1-\frac{1}{\Phi_{-}^{2}(0)}\right]}},
\end{equation}
by considering $\Phi_{-}(0)>0$. It implies to set $\Phi_{-}^{2}(0)-\Lambda_{-}(0)\geqslant1$ to guarantee $|z_{-}(T)|\geqslant0$.  If $\Phi_{-}(0)\gg1$, we can neglect terms of order $\mathcal{O}\left[1/\Phi_{-}^{2}(0)\right]$, and then Eq. \eqref{zmax} reduces to
\begin{equation}
|z_{-}(T)| \approx 1.
\end{equation}
Just as expected under these considerations, we note that  both times $T_{\pm}$ becomes approximately equal to $T$. In other words, we have the approximation
\begin{equation}
\label{int3}
T \approx T_{\pm} \approx \frac{1}{2}\sqrt{\ln\left[\Phi_{-}(0) - \frac{\Lambda_{-}(0)}{\Phi_{-}(0)}\right]}.
\end{equation}
Then, we numerically check the previous approximations to $|z_{-}(T)|$ in Fig. \ref{fig:z-}, by plotting the modulus of the free parameter $|z_{-}(t)|$ in the dimensionless time scale $\gamma t$ with $\gamma=1/2\, \mathtt{s}^{-1}$ assuming the initial values for the Dyson map parameters $\Phi_{-}(0) = 10^{2}$ and $\Lambda_{-}(0) = 10^{-2}$. Whereas Fig. \ref{fig:PhiLambda-} shows the Dyson map parameters $\Phi_{-}(t)$ and $\Lambda_{-}(t)$ in the dimensionless time scale $\gamma t$ assuming the same parameters set in Fig. \ref{fig:z-}. As the time goes to $T$, we have  $\Phi_{-}(T)\gtrsim1$ and $\Lambda_{-}(T)\gtrsim0$.
At the initial time, the Dyson map parameters in Eq. \eqref{TDDM-Ks} assume the values $\epsilon_{-}(0)\approx 11.52$ and $\mu_{-}(0)\approx0.12\,e^{i\varphi_{-}(0)}$. Notice that both parameters are needed to engineering the initial state $|\Psi_{-}(0)\rangle=\hat{\eta}_{-}^{-1}(0)|0,0\rangle$ necessary to achieve the infinite squeezing. Additionally, in face of our assumptions, $\Phi_{-}(0)\gg 1$, our results are in completely agreement to the analysis done in Ref. \cite{deponte2019, deponte2021E, dourado2021}.

\begin{figure}[h!]
\includegraphics[width=\linewidth]{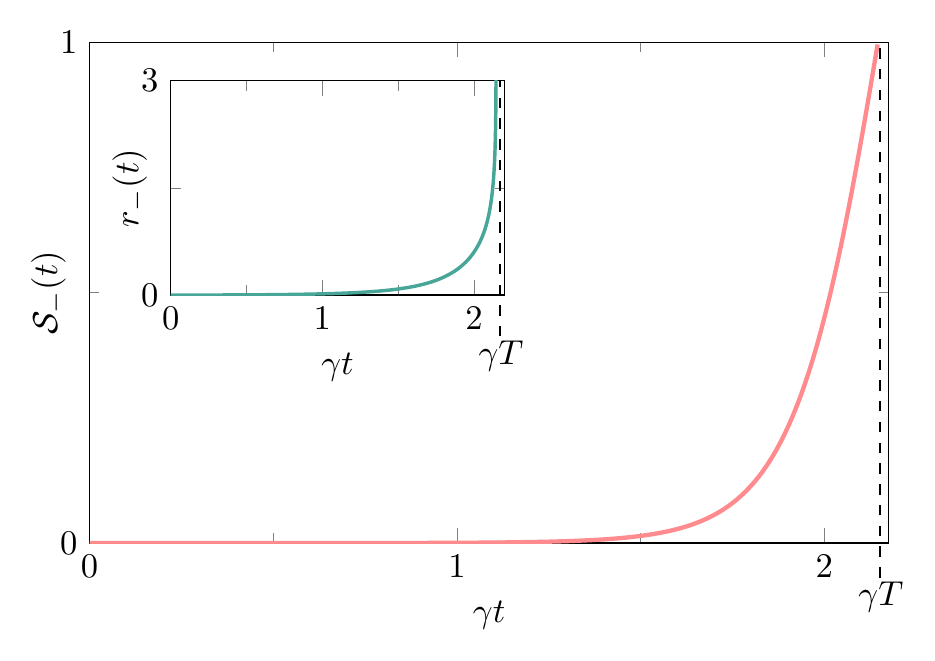}
\caption{\label{fig:S-} The linear entropy $\mathcal{S}_{-}(t)$ against the dimensionless time $\gamma t$ with $\gamma=1/2\,\mathtt{s}^{-1}$ for the initial values $\Phi_{-}(0) = 10^{2}$ and $\Lambda_{-}(0) = 10^{-2}$. Also, we plot the degree of squeezing $r_{-}(t)$ in the inset. The critical dimensionless time at which the degree of squeezing diverges is given by $\gamma T \approx 2.15 $.   }
\end{figure}

Furthermore, as mentioned before, in according to Eq. \eqref{Sl} the maximum entanglement occurs in the limit $r_{-}(t) \rightarrow \infty$, and therefore, from Eq. \eqref{rSol} with $l=1$, we see this happens exactly at $t=T_{+}$ expressed in Eq. \eqref{int2}. Nevertheless, $T_{+}$ can be approximated to $T$ as given by Eq. \eqref{int3} by considering $\Phi_{-}(0)\gg1$. In Fig. \ref{fig:S-}, we plot the linear entropy \eqref{Sl} in the dimensionless time scale $\gamma t$ assuming the same parameters of the earlier plots. Additionally, we also plot the degree of squeezing $r_{-}(t)$ in the inset at Fig. \ref{fig:S-}, in which we observe the further squeeze the state is, the greater entanglement measure between modes becomes. In our case, the degree of squeezing tends towards the infinite, and the linear entropy approximates to its maximum value which corresponds to the unity.

\subsection{$\mathfrak{su}(2)$ entanglement}
The $\mathfrak{su}(2)$ Lie algebra can be expressed in terms of two bosonic operators as follows \cite{wodkiewicz1985,buzek1989}: 
\begin{subequations}
        \begin{align}
        \label{2msu2}
        \hat{K}_{0} &= \frac{1}{2}(  \hat{a}_{1}^{\dagger}\hat{a}_{1} - \hat{a}_{2}^{\dagger}\hat{a}_{2}),
        \\
        \hat{K}_{+}&=\hat{K}_{-}^{\dagger} = \hat{a}_{1}^{\dagger}\hat{a}_{2},
        \end{align}
\end{subequations}
which is nothing but Schwinger's representation of angular momentum.
In the unitary representation of the Lie algebra previously assumed, the total number $n$ of bosons in the system is constant. The vacuum state is understood as the state with no boson in mode $1$ and $n = 2j$ bosons in mode $2$ with $j=0,\frac{1}{2},1,\frac{3}{2},\cdots$ meaning that $\vert\psi_{+}(0)\rangle = \vert 0, n \rangle$, such that $\hat{K}_{-}\vert 0, n \rangle = 0$, for which we have assumed $r_{+}(0)=0$. Under these conditions, the time-evolved state in Eq. \eqref{state-t} reduces to
\begin{equation}
\!\!\!\!\!\vert\psi_{+}(t)\rangle =\zeta_{n}
 \cos^{n}{r_{+}}\sum_{k=0}^{n}C_{n,k}^{\frac{1}{2}}e^{ik\phi_{+}}\tan^{k} r_{+}\vert k, n-k \rangle,
\end{equation}
where $\zeta_{n}=\exp \big [ i n \tilde{\Omega}_{+} \big ]$ is a time-dependent global phase factor while $C_{n,k} = {n \choose k} $ the binomial coefficient.

\begin{figure}[h!]
\includegraphics[width=\linewidth]{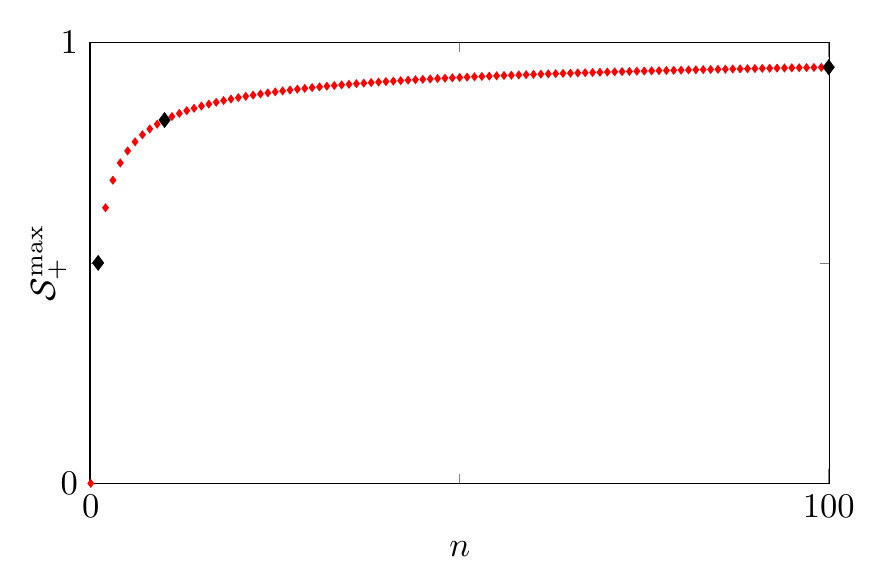}
\caption{\label{fig:Sn} The linear entropy in function of the total number $n$ of bosons when $r_{+}=\pi/4$. Each point corresponds to the maximum entanglement measure for a given $n$. In this case, the linear entropy approximates to unity in the limit $n\rightarrow\infty$. The black diamonds are use to indicate the cases plotted as function of time in Fig. \ref{fig:S+}.}
\end{figure}

Similarly as done before for the case of the $SU(1,1)$,  we start from the reduced density matrix for the first mode written as
\begin{equation}
\hat{\rho}_{+}^{(1)}(t) = \sum_{k=0}^{n}\frac{C_{n,k} \tan^{2k} r_{+}}{[1+\tan^{2}{r_{+}}]^{n}}\vert k \rangle \langle k \vert,
\end{equation}
for which the correspondent linear entropy $\mathcal{S}_{+}(t)$ is given by
\begin{equation}
\mathcal{S}_{+}(t) = 1 - \cos^{n}(2r_{+}) \mathrm{P}_{n}{\left[\frac{1+\cos^{2}(2 r_{+})}{2\cos(2 r_{+})}\right]},
\label{Ss}
\end{equation}
determined in terms of the Legendre Polynomial $\mathrm{P}_{n}(\cdot)$. Note that while for the $SU(1,1)$ the linear entropy depends on the hyperbolic functions on $r_{-}$, for the case of $SU(2)$ the dependence occurs in terms of polynomial forms on the trigonometric functions, evidencing the periodic behavior of the entropy in time.
Moreover, it can be verified that, for a given $n$, the maximum entanglement measure occurs for $r_{+}\rightarrow\pi/4 + 2k\pi$ with $k\in\mathbb{Z}$. At this limit, the linear entropy \eqref{Ss} reduces to
\begin{equation}
\mathcal{S}_{+}^{\text{max}} = 1 - \frac{\Gamma\left(n+\frac{1}{2}\right)}{\sqrt{\pi}n!}.
\label{Slim}
\end{equation}
We see the entanglement measure has a direct dependence on the total number $n$ of bosons, and it goes to the maximum value $\mathcal{S}_{+}^{\text{max}}=1$ when $n\rightarrow\infty$. This dependence on $n$ is plotted in the Fig. \ref{fig:Sn}.

Here, we also restrict our analysis to the linear imaginary part $\omega_{\text{I}}(t)$ given by Eq. \eqref{omega}. Thus, before analyzing the possibility of achieving the maximum entanglement as done for the $SU(1,1)$ case, we investigate the Dyson map parameters.

\begin{figure}[b!]
        \includegraphics[width=\linewidth]{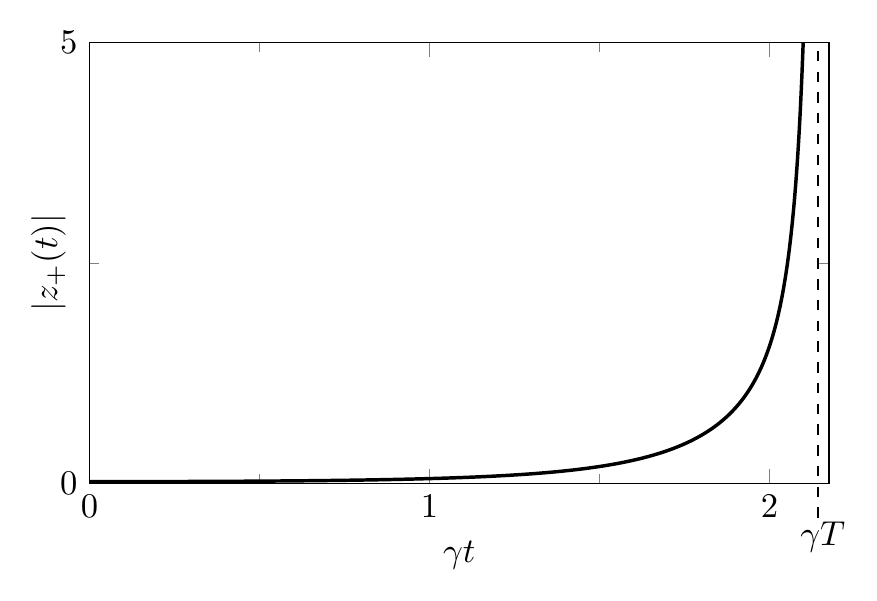}
        \caption{\label{fig:z+} The modulus of free Dyson map parameter $|z_{+}(t)|$ in the dimensionless time $\gamma t$ with $\gamma=1/2\, \mathtt{s}^{-1}$ for the initial values $\Phi_{+}(0) = 10^{2}$ and $\Lambda_{+}(0) = 10^{-2}$. We evaluate the time-evolution until the dimensionless time $\gamma T \approx 2.15 $.}
\end{figure}

From Eq. \eqref{z}, we can obtain the time at which the modulus of the free parameter goes to its upper limit, which corresponds to $|z_{+}(t)|\rightarrow\infty$. We have this happens when the following condition is complied:
\begin{equation}
\Phi_{+}^{2}(T) = 1-\Lambda_{+}(T),
\end{equation}
for $0<\Lambda_{+}(T)\leq1$, where we are supposing it is achieved at the time instant $T$, which is given by
\begin{equation}
\label{int5}
T = \frac{1}{\gamma} \sqrt{\frac{1}{2}\ln\left[\Phi_{+}^{2}(0)\frac{\left(1+\frac{\Lambda_{+}(0)}{\Phi_{+}^{2}(0)}\right)^{2}+\frac{1}{\Phi_{+}^{2}(0)}}{1+\frac{1}{\Phi_{+}^{2}(0)}}\right]}.
\end{equation}
In fact, $T$ represents the largest time in which the Dyson map \eqref{TDDM-Ks} can hermitize the non-Hermitian Hamiltonian through the relation expressed in Eq. \eqref{h0}. The behavior of the free Dyson map parameter $|z_{+}(t)|$ is represented in the Fig. \ref{fig:z+}.

We can suppose the function $r_{+}(t)\rightarrow\pi/4$ at the time $t=T^{\prime}$, such that
\begin{equation*}
\lim_{t\rightarrow T^{\prime}} r_{+}(t) = \frac{\pi}{4},
\label{is}
\end{equation*}
which is achieved when the following equation holds
\begin{equation}
\label{int4}
T^{\prime} = \frac{1}{\gamma}\sqrt{\ln\left[\Phi_{+}(0)\frac{1+\frac{1}{\Phi_{+}(0)}+\frac{\Lambda_{+}(0)}{\Phi_{+}^{2}(0)}}{1-\frac{1}{\Phi_{+}^{2}(0)}}\right]},
\end{equation}
by assuming $r_{+}(T)=\pi/4$ and $l=1$ in Eq. \eqref{rSol}. Additionally, if we consider $\Phi_{+}\gg1$ and $\Lambda_{+}\ll1$, we can neglect terms of order $\mathcal{O}\left[1/\Phi_{-}(0)\right]$. So that, under these assumptions, both times $T$ and $T^{\prime}$ expressed into Eqs. \eqref{int5}  and \eqref{int4}, respectively, become approximately equal to each other, and then
\begin{equation}
\label{Tsu2}
T\approx T^{\prime}\approx \frac{1}{\gamma}\sqrt{\ln\Phi_{+}(0)}.
\end{equation}%
\begin{figure}[t!]
\includegraphics[width=\linewidth]{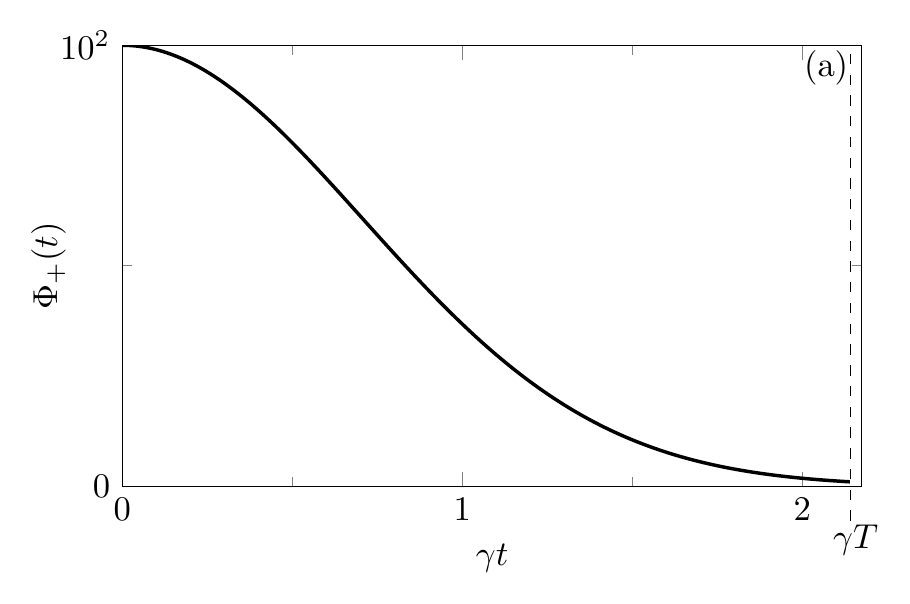}
\includegraphics[width=\linewidth]{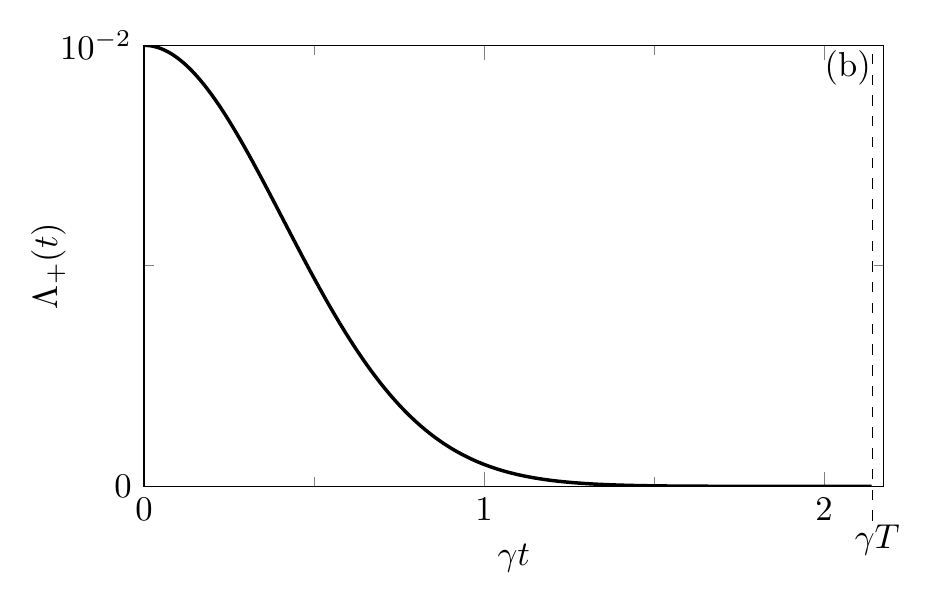}
\caption{\label{fig:PhiLambda+} The Dyson map parameters (a) $\Phi_{+}(t)$ and (b) $\Lambda_{+}(t)$  against the dimensionless time $\gamma t$ with $\gamma=1/2\, \mathtt{s}^{-1}$ for the initial values $\Phi_{+}(0) = 10^{2}$ and $\Lambda_{+}(0) = 10^{-2}$. We evaluate the time-evolution until the dimensionless time $\gamma T \approx 2.15 $.}
\end{figure}

Therefore, there are no inconsistencies occurring in time-evolution of the Dyson map parameters as can be seen numerically in the Fig. \ref{fig:PhiLambda+}. For these plots, we consider a dimensionless time scale $\gamma t$ with $\gamma=1/2\,\mathtt{s}^{-1}$, and the initial time Dyson map parameters $\Phi_{+}(0)=10^{2}$ and $\Lambda_{+}(0)=10^{-2}$. Moreover, the time-dependent Dyson map parameters $\epsilon_{+}(t)$ and $|\mu_{+}(t)|=\epsilon_{+}(t)|z_{+}(t)|/2$ can be obtained from Eqs. \eqref{epsilon} and \eqref{z}.  Both parameters are needed to engineering the initial state $|\Psi_{+}(0)\rangle=\hat{\eta}_{+}^{-1}(0)|0,n \rangle$ necessary to achieve the maximum entanglement value. Thus, at the initial time, the parameters of Dyson map $\epsilon_{+}(0)\approx 11.51$ and $\mu_{+}(0)\approx 0.12\,e^{i\varphi_{+}(0)}$.

\begin{figure}[h!]
\includegraphics[width=\linewidth]{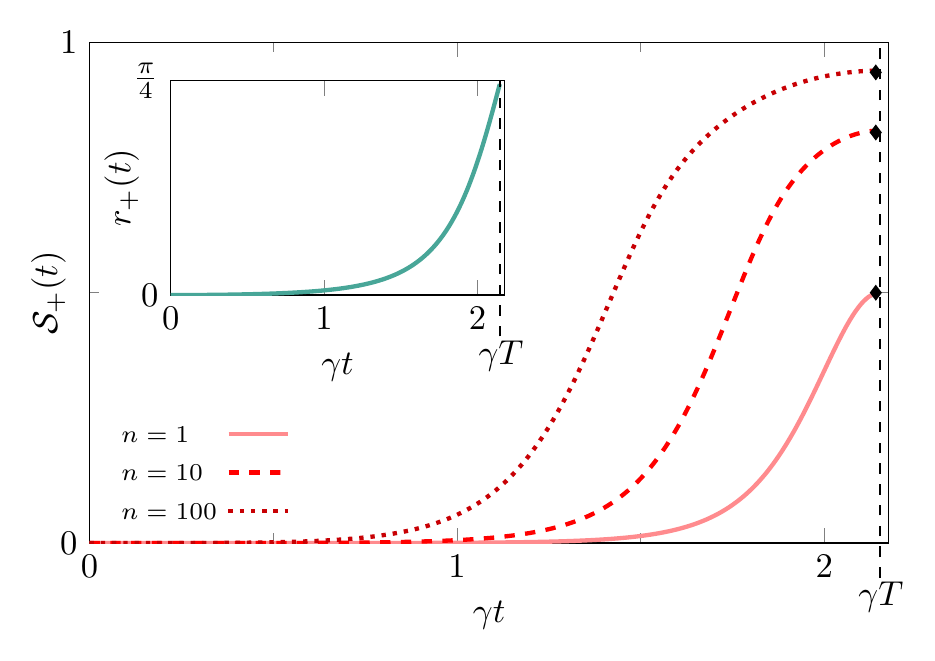}
\caption{\label{fig:S+} For $n=1$ (solid line), $n=10$ (dashed line) and $n=100$ (dotted line), we plot the linear entropy $\mathcal{S}_{+}(t)$ against the dimensionless time $\gamma t $ with $\gamma=1/2\, \mathtt{s}^{-1}$ for the initial values $\Phi_{+}(0) = 10^{2}$ and $\Lambda_{+}(0) = 10^{-2}$. The dimensionless time at which the maximum entanglement occurs is  $\gamma T \approx 2.15$. The black diamonds on the maximum values of entropy are used to indicate the correspondent value at the plot in Fig. \ref{fig:Sn} by the same symbols.}
\end{figure}

Furthermore, at the time $T$ given by Eq. \eqref{Tsu2},  $r_{+}(T) \approx \pi/4$ under the previous approximations. This value corresponds to the maximum entanglement measure for a given $n$, expressed in Eq. \eqref{Ss}. The behavior of the linear entropy \eqref{Ss} is plotted in the dimensionless time scale $\gamma t $ with $\gamma=1/2\mathtt{s}^{-1}$ in Fig. \ref{fig:S+}, together with the parameter $r_{+}(t)$ in the inset. We observe the further $r_{+}$ approximates to $\pi/4$, more the linear entropy goes to its maximum value dependent on $n$, as expressed in Eq. \eqref{Slim} and illustrated in Fig. \ref{fig:Sn}. We indicate by big diamonds in Fig. \ref{fig:Sn}, the correspondent cases analyzed in Fig. \ref{fig:S+}.

\subsection{Towards a generalization: multimode realizations}
Our analysis can be generalized to the multimode bosonic realizations of both Lie algebras. For instance, we can  consider the multimode bosonic realizations proposed by Lo and coworkers in Ref. \cite{Lo1993,sollie1993,Lo1994,Lo1995}. It is clear that the mapping between non-Hermitian and Hermitian representations through the time-dependent Dyson map remains the same as discussed until Section \ref{sec:K0}, see Eqs. \eqref{k0nh} and \eqref{k0h}. From Refs. \cite{Lo1993,sollie1993,Lo1994,Lo1995}, the $\mathfrak{su}(1,1)$ multimode bosonic realization reads as
\begin{subequations}
        \begin{align*}
        \hat{K}_{0}	&=\sum_{i,j=1}^{N}\mathsf{K}_{ij}\left(2\hat{a}_{i}^{\dagger}\hat{a}_{j}+\delta_{ij}\right),
        \\
        \hat{K}_{+}	&=\hat{K}_{-}^{\dagger} = \frac{1}{2}\sum_{i,j=1}^{N}\mathsf{k}_{ij}\hat{a}_{i}^{\dagger}\hat{a}_{j}^{\dagger},
        \end{align*}
\end{subequations}
for which the equalities $\mathsf{K}_{ij}=\mathsf{K}_{ji}^{\ast}$ and $\mathsf{k}_{ij}=\mathsf{k}_{ji}$ are verified. They arise from the commutation relations in Eq. \eqref{LieAlgebras} with $\mathfrak{s}=-1$ that require
\begin{subequations}
        \label{su11_rep}
        \begin{align*}
        \mathsf{K}_{ij} &= \frac{1}{4}\sum_{k=1}^{N} \mathsf{k}_{ik}\mathsf{k}_{kj}^{\ast}
        ,\\
        \mathsf{k}_{ij} &=\sum_{k,l=1}^{N} \mathsf{k}_{ik}\mathsf{k}_{kl}^{\ast}\mathsf{k}_{lj}.
        \end{align*}
\end{subequations}
In addition, the $\mathfrak{su}(2)$ multimode bosonic realization can be written as
\begin{subequations}
        \begin{align*}
        \hat{K}_{0}	&=\sum_{i,j=1}^{N}\mathsf{J}_{ij}\hat{a}_{i}^{\dagger}\hat{a}_{j},
        \\
        \hat{K}_{+}	&=\hat{K}_{-}^{\dagger} = \sum_{i,j=1}^{N}\vartheta_{ij}\hat{a}_{i}^{\dagger}\hat{a}_{j}.
        \end{align*}
\end{subequations}
Due to the commutation relations  in \eqref{LieAlgebras} with $\mathfrak{s}=1$, the following
conditions must hold
\begin{subequations}
        \label{su2_rep}
        \begin{align*}
        \mathsf{J}_{ij} &= \frac{1}{2}\sum_{k=1}^{N} \left(\vartheta_{ik}\vartheta_{jk}^{\ast} - \vartheta_{ki}^{\ast}\vartheta_{kj}\right),
        \\
        \vartheta_{ij}  &=\sum_{k=1}^{N} \left(\mathsf{J}_{ik}\vartheta_{kj} - \vartheta_{ik}\mathsf{J}_{kj}\right).
        \end{align*}
\end{subequations}
Therefore, we can develop a similar analysis to characterize entanglement between uncoupled modes by setting the representations in which the $\hat{K}_{0}$ operator is composed only by uncoupled modes $\hat{K}_{0}\propto\hat{a}_{k}^{\dagger}\hat{a}_{k}^{}$, while the $\hat{K}_{+}=\hat{K}_{-}^{\dagger} \propto \hat{a}_{j}^{\dagger}\hat{a}_{k}^{\dagger}$ or $\hat{K}_{+}	=\hat{K}_{-}^{\dagger} \propto \hat{a}_{j}^{\dagger}\hat{a}_{k}^{}$ with $j\neq k$ for the $\mathfrak{su}(1,1)$ and $\mathfrak{su}(2)$, respectively.
The computation of the multipartite entanglement measure is not an easy
task, although it is possible to investigate the existence of
entanglement through multipartite entanglement criteria as the proposed
by Hillery et al. in Ref. \cite{hillery2010}. Also, bipartite multimode entanglement measures can be analyzed by means of the linear entropy of each bipartition.

\section{Conclusion}
\label{sec:conclusion}

In summary, we obtained the explicit solutions for Dyson map and time-evolution operator without mention the Lie Algebras realization. In what follows, we note that there is no apparent interaction between the modes in the non-Hermitian Hamiltonian \eqref{k0nh}, when we look from the point of view of conventional quantum mechanics and its trivial Hilbert space metric.  Nevertheless, the hermitization procedure applied to \eqref{k0nh} consists in defining a non-trivial dynamical metric that leads to a Hermitian counterpart \eqref{k0h}, in which the interaction between the two modes becomes evident, and quantum correlations such as entanglement can exist in this case. The key point in this discussion is that non-Hermitian Hamiltonian operators describing a non-interacting two-modes system can induce quantum correlations such as entanglement between the modes due to non-hermiticity only.

Basically, we presented a time-dependent non-Hermitian Hamiltonian embedding the generators of $SU(1,1)$ and $SU(2)$ Lie groups in a unified. By applying the Ansatz of time-dependent Dyson map and metric operator described by the same algebraic structure, we verify that the Hermitian counterpart also exhibits a $SU(1,1)$ or $SU(2)$ dynamical symmetries. So that, the Hermitian counterpart becomes independent of the algebra realization, as demonstrated in Ref. \cite{quesne2007} for the time-independent Swanson model. Furthermore, the obtained Hermitian counterpart \eqref{h} reduces to that one studied in Refs. \cite{maamache2017,koussa2018} by setting $\mathcal{U}_{\mathfrak{s}}=\mathcal{V}_{\mathfrak{s}}=0$ rather than the general assumptions read in \eqref{constraint}. Nevertheless, these more general constraints have allowed us to derive a non-trivial result, from which a non-Hermitian system with a non-apparent interacting term exhibits entanglement, which is verified by mapping on its Hermitian counterpart. In terms of two-modes Lie algebra realizations, we show that the uncoupled non-Hermitian Hamiltonian \eqref{k0nh} has a Hermitian counterpart \eqref{k0h} in which the modes are coupled, and it leads to a non-trivial entanglement only due to the time-dependent complex frequency: which may be seen as a metric-dependent entanglement. The nontrivial dynamical Hilbert space metric allows us to correlate quantum systems even in absence of interaction terms in non-Hermitian Hamiltonian. It happens due to the generality of the Hermitian Dyson map structure \eqref{TDDM-Ks}, which makes the metric to be non-local depending on the choice of parameters. We investigated, for both Lie algebras, the case where the non-hermiticity is given by a frequency $\omega(t) = \omega_{\text{R}}(t) + i \gamma^2 t$, also considering the transformation $\Phi_{-}\rightarrow-\Phi_{-}$ for comparison to the results obtained in Ref. \cite{dourado2021}. Our investigations showed that the maximum entanglement measure by means of linear entropy is achieved at a finite time interval $T$ given by Eqs. \eqref{int3} and \eqref{Tsu2}, from which we can write $T \approx (1/\gamma)\sqrt{\ln{\Phi_{\pm}}}$ by assuming $\Phi_{\pm}\gg1$, and neglecting the terms proportional to $\Lambda_{\pm}/\Phi_{\pm}\ll1$.

Although the authors in Ref. \cite{scolarici2009} argued that descriptions of two interacting subsystems are possible if and only if the metric operator of the compound system can be obtained as a tensor product of positive operators on component spaces.  We believe this statement is too restrictive and not necessary since the Dyson map is, in general, not unique and it may lead to a wide class of non-trivial metrics associated with the Hilbert space. Furthermore, the interpretation of non-Hermitian quantum systems (with nontrivial metric operators in their Hilbert spaces) is made by mapping the problem to locally Hermitian ones with a standard trivial metric, which allows a clear description of dynamics.

Thus, the algebraic structure of quantum mechanics allows us to go towards generalizations of many interesting mathematical structures and the physical phenomena associated with them. In what concerns non-Hermitian quantum mechanics, the algebraic language seems to play a key role in understanding the physical aspects of non-Hermitian physics, once the hermiticity is closely related to the geometry of Hilbert space encoded in its metric.  For instance, in Ref. \cite{uhdre2022} was shown that a deformed algebra applied to the study of Dirac oscillator leads to a natural map of the relativistic system in the non-Hermitian version of the well-known Jaynes-Cummings optical model. 
The symmetrical approaches may be useful tools to understand non-Hermitian effects which are naturally explained by a duality between non-Hermitian models in flat spaces and their counterparts, which could be Hermitian, in curved spaces \cite{lv2022}. In this regard, the $SU(1,1)$ and $SU(2)$ Lie groups and their correspondent algebras may provide future investigation towards hyperbolic and spherical spaces \cite{kishimoto2001}, and their non-stationary generalizations.

In conclusion, our work may contribute to the theoretical progress of time-dependent non-Hermitian quantum systems in the context of compound systems to bring new possibilities of applications in quantum information areas, many-body quantum physics, and also for improving our understanding of the mathematical structure of quantum mechanics. Although engineering effective non-Hermitian Hamiltonians seems to be a feasible task by considering continuous measurements and post-selection \cite{gopalakrishnan2021}, or even through adiabatic elimination techniques \cite{gamel2010,james2007}, we believe the most intriguing non-Hermitian phenomena in closed quantum systems are encoded at the nontrivial Hilbert space metrics, and engineering them is still a challenge. 
Recent discussions about curving Hilbert space as done in Refs. \cite{lv2022} also appoint to this fact. Perhaps, the experimental breakthroughs of curved spaces in nanophotonic structures \cite{bekenstein2017,kollar2019}, in which curved spaces may be designed, might shed light on future investigations in this subject to pave the way on build nontrivial geometries embracing the non-Hermitian physics.


\section*{Acknowledgments}
The authors are grateful to Miled H.Y. Moussa and R. Auccaise for helpful
discussions. This work was partially supported by the Brazilian agencies
Conselho Nacional de Desenvolvimento Cient\'ifico e Te\-cnol\'ogico
(CNPq), and Instituto Nacional de Ci\^{e}ncia e Tecnologia de
Informa\c{c}\~{a}o Qu\^{a}ntica (INCT-IQ).  It was also financed by the
Co\-or\-dena\c{c}\~{a}o de Aperfei\c{c}oamento de Pessoal de N\'{i}vel
Superior (CAPES, Finance Code 001). DC thanks to Instituto Serrapilheira and Barbara Amaral. FMA also acknowledges CNPq Grant no. 314594/2020-5.

\bibliographystyle{apsrev4-2}
\bibliography{references.bib}

\end{document}